\documentclass[a4paper,aps,pre,twocolumn,showpacs,showkeys,floatfix]{revtex4}
\usepackage{graphicx}

\newcommand{\middlefig}{.23\textwidth}
\newcommand{\singlefig}{.38\textwidth}

\begin{document}

\title{Discrete soliton collisions in a waveguide array with saturable
  nonlinearity}

\author{J. Cuevas} \email{jcuevas@us.es}
\affiliation{Grupo de F\'{\i}sica No Lineal.  Departamento de
  F\'{\i}sica Aplicada I., EU Polit\'ecnica.  Universidad de Sevilla,
  C/ Virgen de \'Africa, 7, 41011-Sevilla, Spain}

\author{J. C.
  Eilbeck} \email{J.C.Eilbeck@hw.ac.uk} \affiliation{Department of
  Mathematics, Heriot-Watt University \\Riccarton, Edinburgh, EH14
  4AS, UK }

\date{\today}
\begin{abstract}
  We study the symmetric collisions of two mobile breathers/solitons
  in a model for coupled wave guides with a saturable nonlinearity.
  The saturability allows the existence of breathers with high power.
  Three main regimes are observed: breather fusion, breather
  reflection and breather creation.  The last regime seems to be
  exclusive of systems with a saturable nonlinearity, and has been
  previously observed in continuous models. In some cases a ``symmetry
  breaking'' can be observed, which we show to be an numerical
  artifact.
\end{abstract}



\maketitle

\section{Introduction}

Since the 1960's, a great number of papers have considered the
properties of solitons in nonlinear optic media with a Kerr-type
(cubic) nonlinearity. This media can be modelled by the cubic
Nonlinear Schr\"odinger (NLS) Equation. As it is well known, the NLS
equation is integrable and, in consequence, solitons interact
elastically \cite{ZS72}.

More recently, several authors have studied the properties of solitons
in photo-refractive media \cite{IEEE}. In this case, the equation
describing these media is a modification of the original NLS, which
consists in substituting the Kerr nonlinearity term by another one of
saturable type. This Saturable (SNLS) Equation is nonintegrable and
the soliton collision processes are inelastic, leading to
annihilation, fusion or creation of solitons \cite{SNLS}. This last
phenomena consists of the appearance of three solitons after the
collision of only two of them. Another important feature of the SNLS
is that the behaviour of the solutions is quite generic, being
independent of the details of the mathematical model.

The {\em discrete} version of the NLS equation can be used to describe
nonlinear waveguide arrays within the tight binding approximation
\cite{nature}. The existence and properties of mobile discrete
breathers/solitons in DNLS lattices has been considered in a number of
studies (We use the terms breathers and solitons interchangeably in
this context, also intrinsic localized modes). An early brief study
\cite{ei86} showed that breathers could propagate along the lattice
with a small loss of energy, and could become trapped by
inhomogeneities in the lattice.  Later, a more detailed study
\cite{fe91} suggested that ``exact'' travelling breathers might exist,
at least for some parameter ranges.  The reviews \cite{ej03,krb01}
refer to many other papers in this area.  More recently, work has
concentrated on breathers with infinite oscillating tails
\cite{gff04}, although the question of the {\em existence} of {\em
  exact} breather solutions which tend to zero as $n \rightarrow \pm
\infty$ has not yet been resolved.  Given the long history of mobile
breather solutions of this equation, it is rather surprising that a
systematic study of the collision of two breathers in the DNLS model
has only recently been carried out \cite{PKMF03}. (We mention also
that collisions have been studied in generalised nearly integrable
DNLS model in \cite{DKMF03,CBG97}).

Recently, some studies have considered the existence of mobile
breathers in waveguide arrays in photo-refractive crystal, described
by a DNLS equation with saturable nonlinearity \cite{HMSK04,SKHM04}.
In particular, these papers considered a discrete version of the
Vinetskii-Kukhtarev model \cite{IEEE,VK75}. The key difference
between the cubic DNLS equation and the saturable DNLS equation is
that in the later, the Peierls--Nabarro barrier (the energy
difference between a bond-centred and a site-centred breather with
the same power) is bounded and, in most cases, smaller than in the
former \cite{KC93}. It allows the existence of mobile breathers of
high power.

The aim of the present paper is to study breather-breather collisions
in a saturable DNLS equation and to compare the results with those
obtained in the continuous SNLS and the discrete cubic equation.

\section{Numerical results}

The system we consider is governed by the following equation of motion
\begin{equation}\label{eq:dyn}
    i\dot u_n-\beta\frac{u_n}{1+|u_n|^2}+(u_{n+1}-2u_n+u_{n-1})=0.
\end{equation}
This model has two conserved quantities: the Hamiltonian
$H=\sum_n[\beta\log(1+|u_n|^2)+|u_{n-1}-u_n|^2]$ and the power (or
norm) $P=\sum_n|u_n|^2$.

In order to reduce the dimension of the large parameter space to be
considered, we have fixed $\beta$ to $\beta=2$. Higher values of
$\beta$ lead to solutions that only can be moved for a restricted
set of power values \cite{HMSK04}.

A moving breather $v_n(t)$ is obtained by adding a thrust $q$ to a
stationary breather $u_n$, so that:
\begin{equation}\label{eq:initial_condition}
    v_n(0)=u_n\exp(i\, q\, n).
\end{equation}
Notice that this procedure of obtaining moving breathers is similar to
the marginal mode method introduced in \cite{Aubry,ac98} for Klein--Gordon
lattices.

In the following, we consider the collision of two identical
breathers moving in opposite directions with the same thrust $q$.
Analogously to Ref.\ \cite{PKMF03}, we consider both inter-site (IS)
and on-site (OS) collisions.

\begin{figure}[ht] %fig 2
\begin{center}
\begin{tabular}{c}
    (a)  \\
    \includegraphics[width=\singlefig]{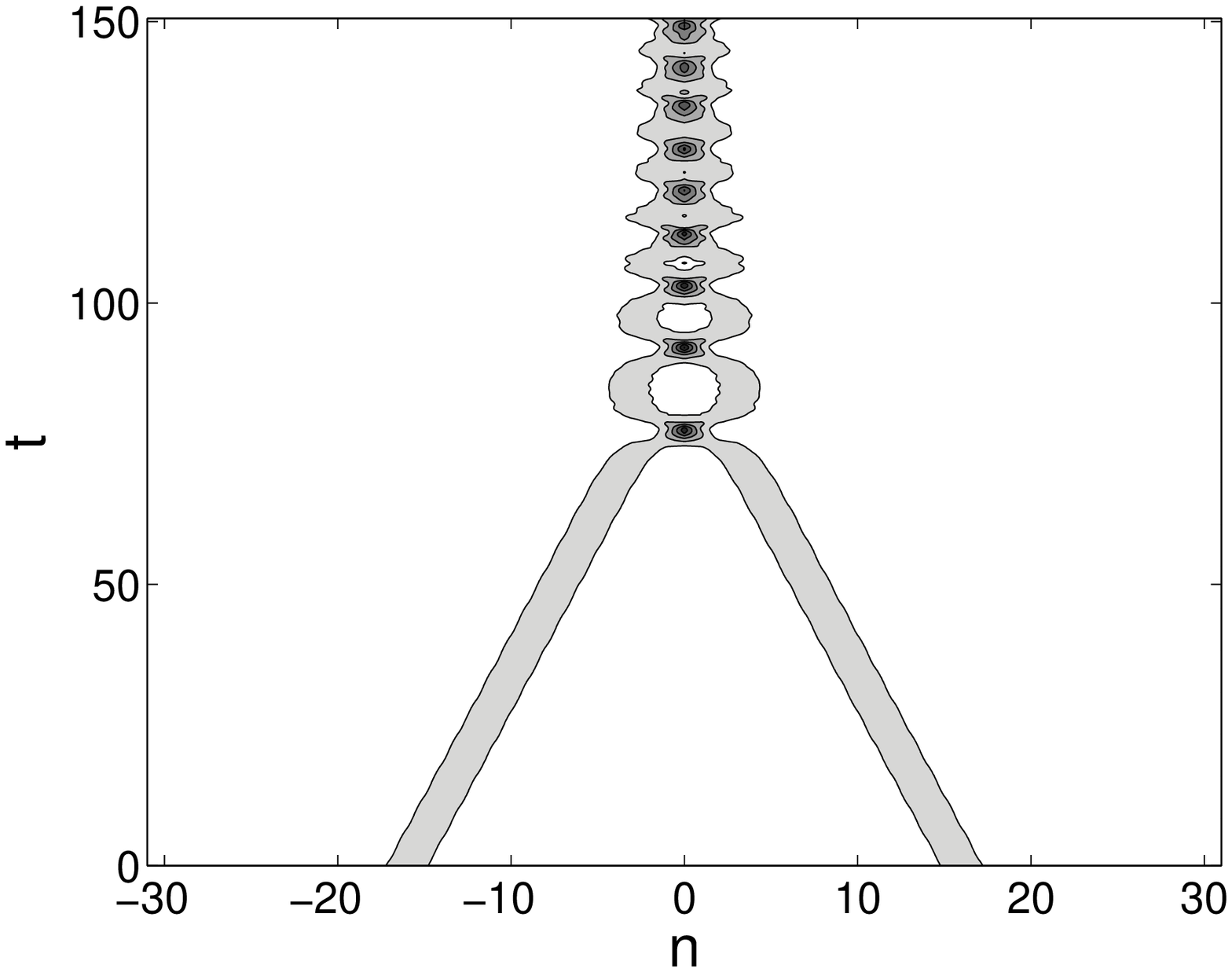} \\
    (b)\\
    \includegraphics[width=\singlefig]{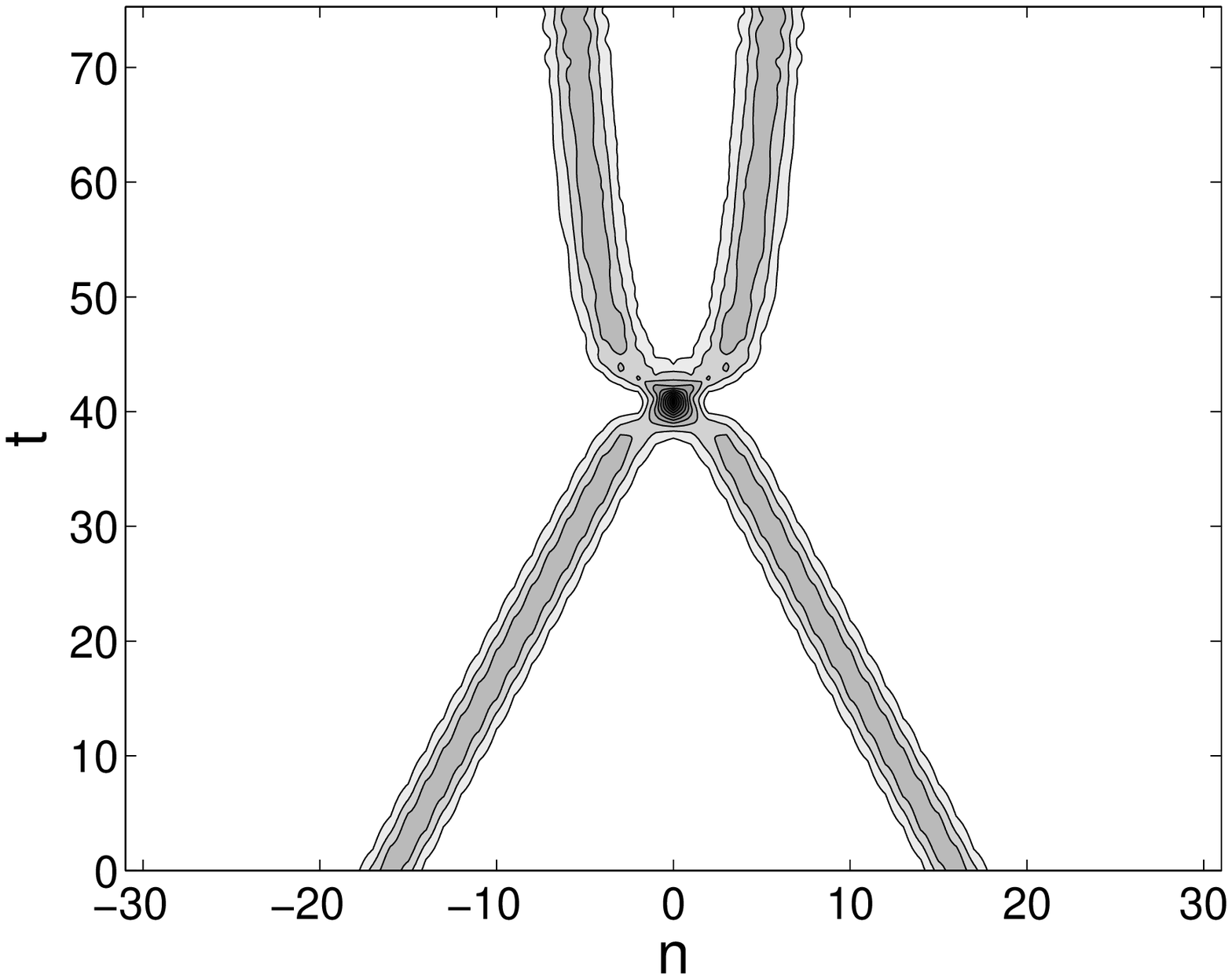}\\
    (c)\\
    \includegraphics[width=\singlefig]{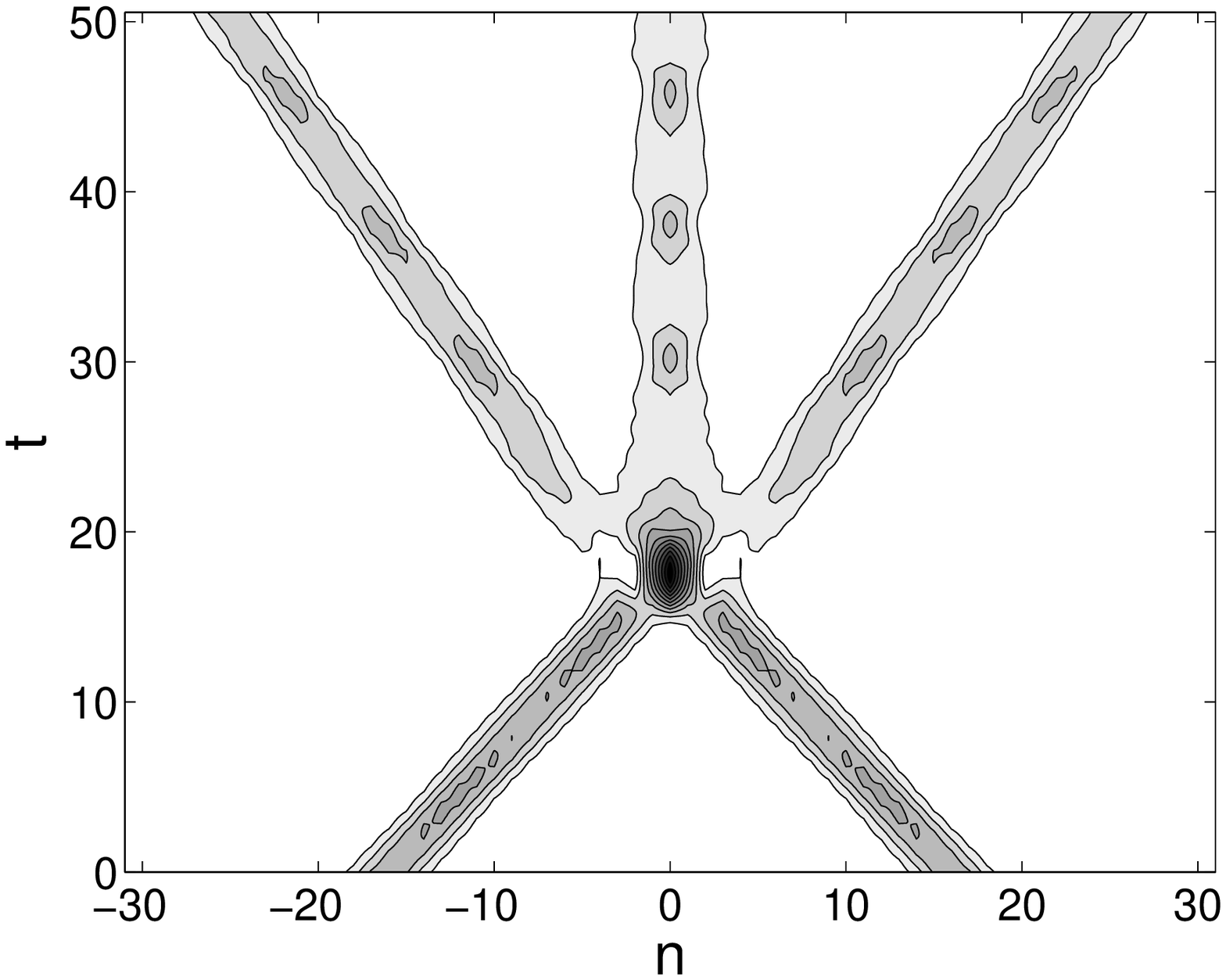}\\
\end{tabular}
\caption{Typical power density plots for (a) bound state formation
  ($P=10$, $q=0.1$), (b) reflection ($P=10$, $q=0.2$), and (c)
  breather creation ($P=70$, $q=0.5$). In all cases, OS collisions are
  considered, although these pictures do not vary considerably for IS
  collisions.}
\label{fig:example}
\end{center}
\end{figure}

The collision scenario we observe for small $P$ is quite
simple: there exists a critical value $q_c$ below which breathers form
a bound state, and above which, breathers are reflected (See
Fig.~\ref{fig:example}a-b for examples of these two cases). It can be
observed that the bound state ``oscillates'' after the collision.  The
amplitude of these oscillations decreases when approaching to the
critical point, whereas their ``period'' increases. (Note that the
``reflection'' case could equally be regarded as a transmission case
as the two breathers are indistinguishable.  In the case of
reflection/transmission, there is some loss of energy of the two
breathers).

\begin{figure}[ht] %fig 3
\begin{center}
\begin{tabular}{cc}
    (a)  & (b)\\
    \includegraphics[width=\middlefig]{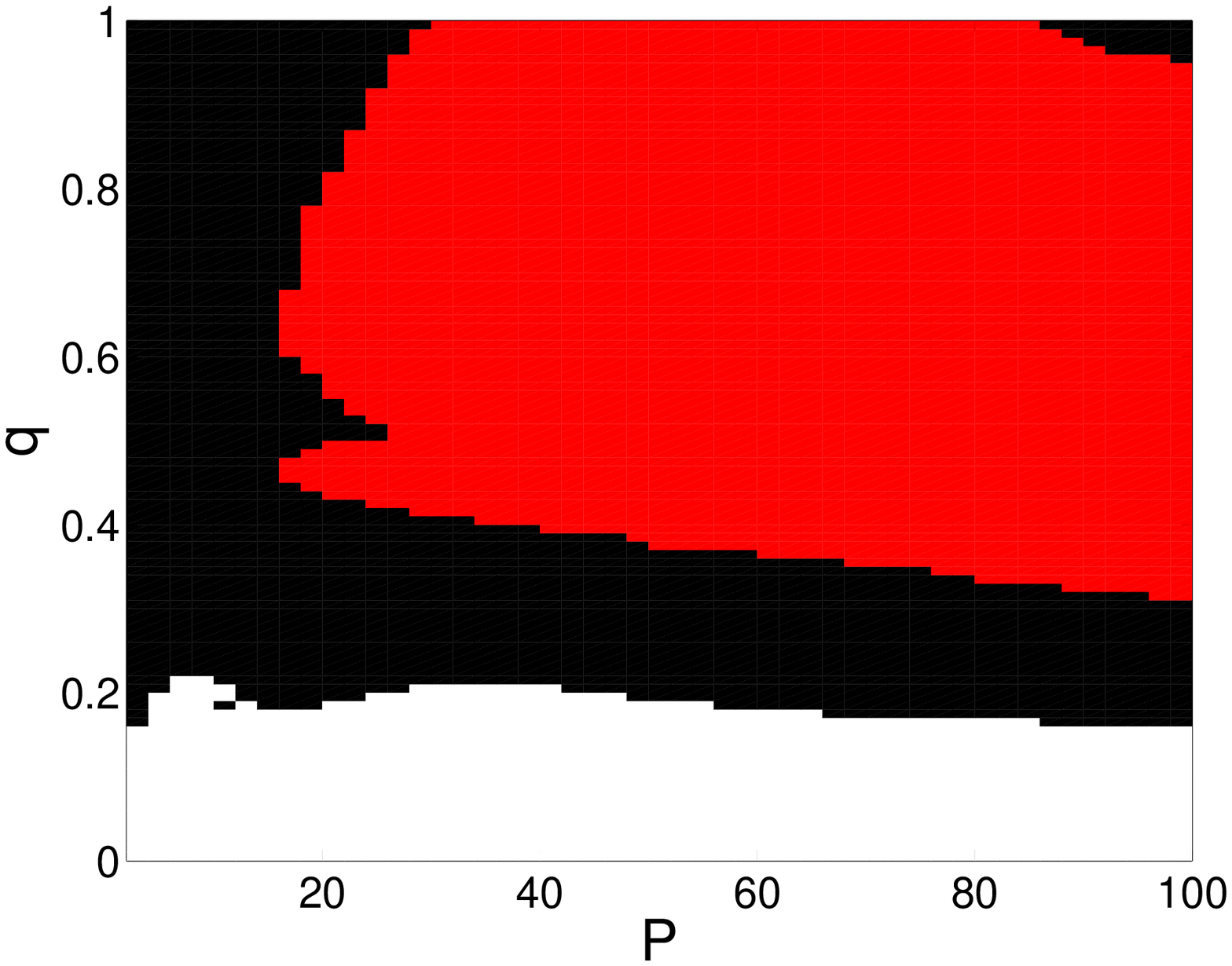} &
    \includegraphics[width=\middlefig]{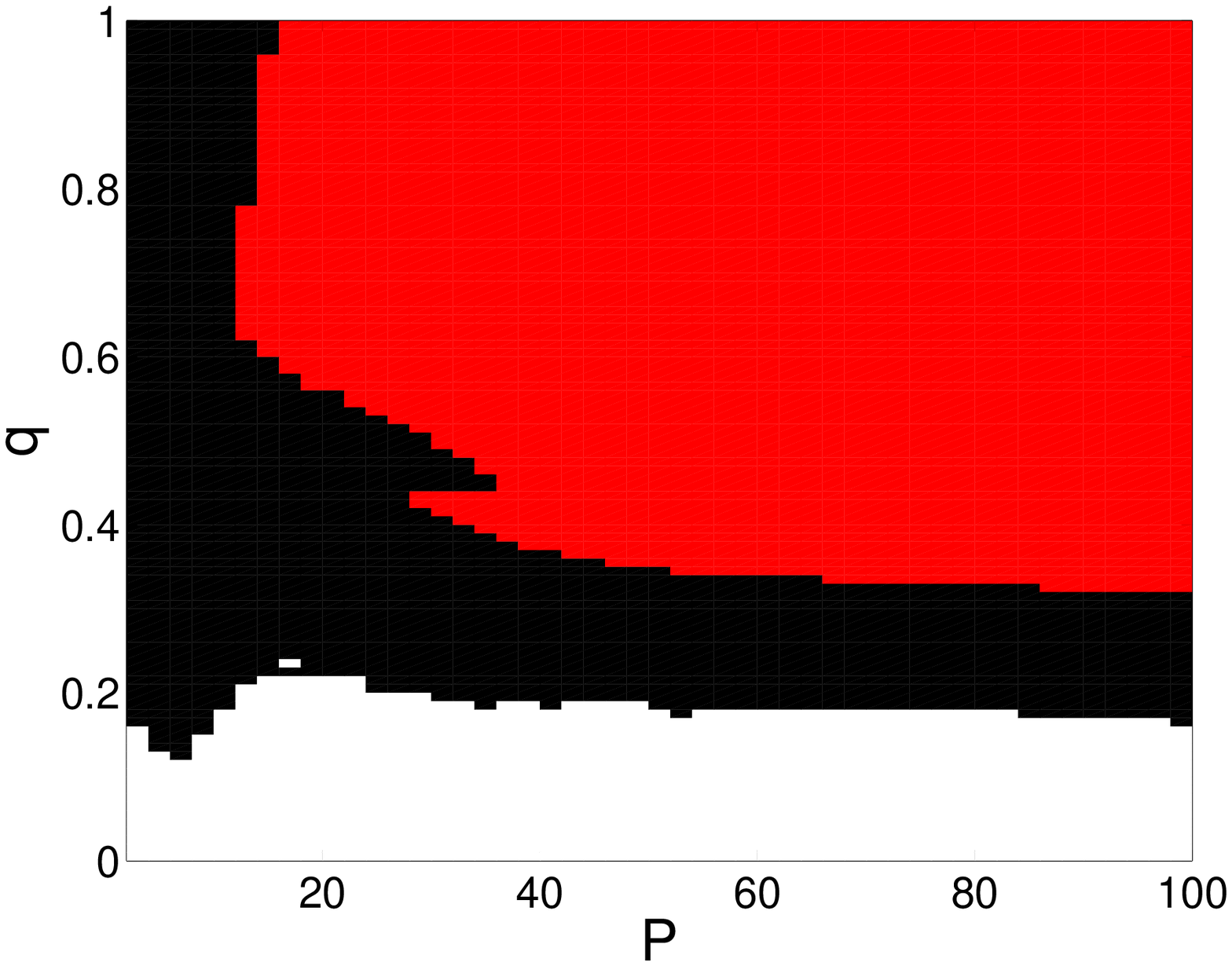}\\
\end{tabular}
\caption{Different regimes observed in (a) IS and (b) OS
  collisions. The colours represent the following: white-merge to a
  single breather; black-reflection; and red-breather creation}%
\label{fig:regimes}
\end{center}
\end{figure}

\begin{figure}[ht] %fig 3
\begin{center}
\begin{tabular}{cc}
    (a)  & (b) \\
    \includegraphics[width=\middlefig]{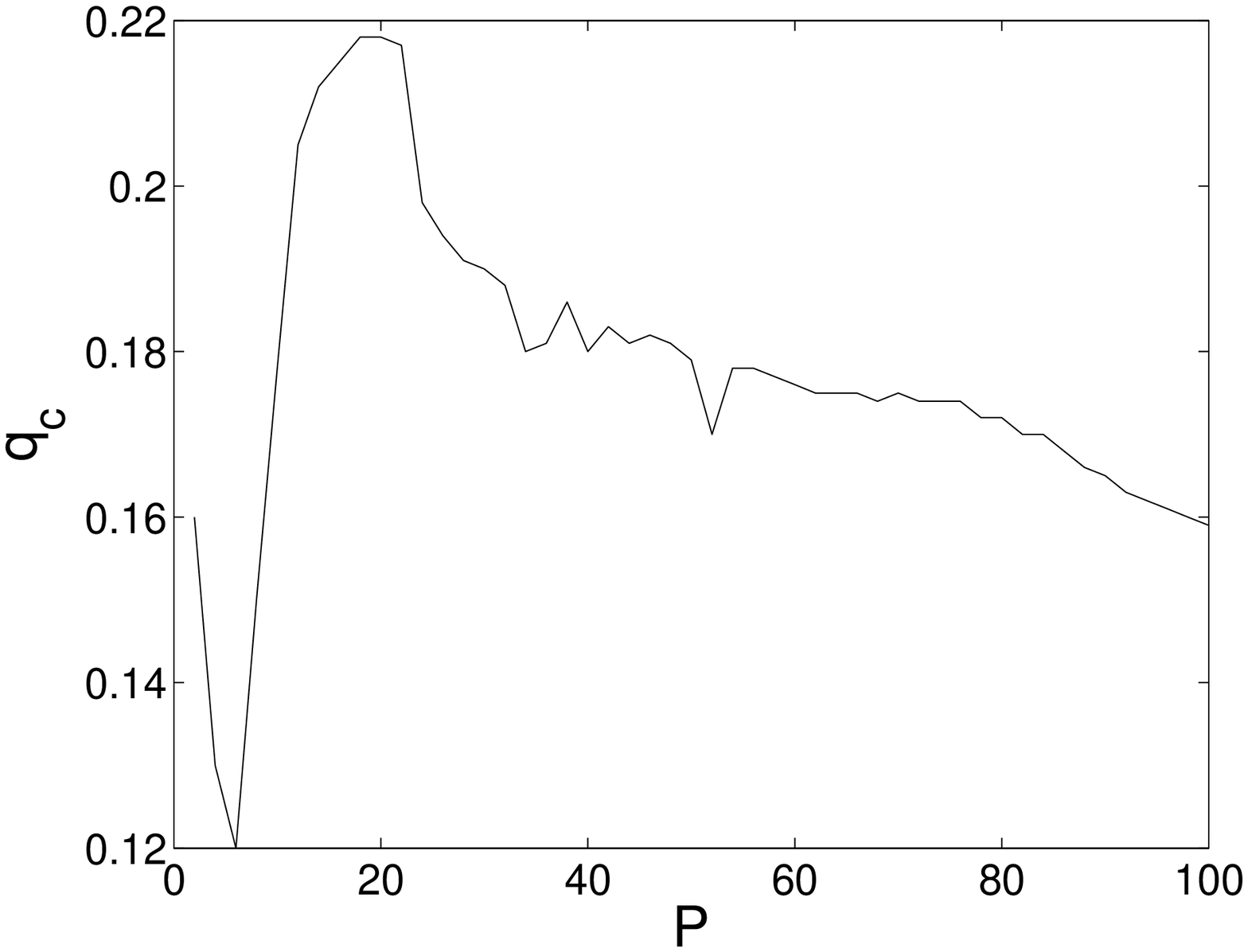} &
    \includegraphics[width=\middlefig]{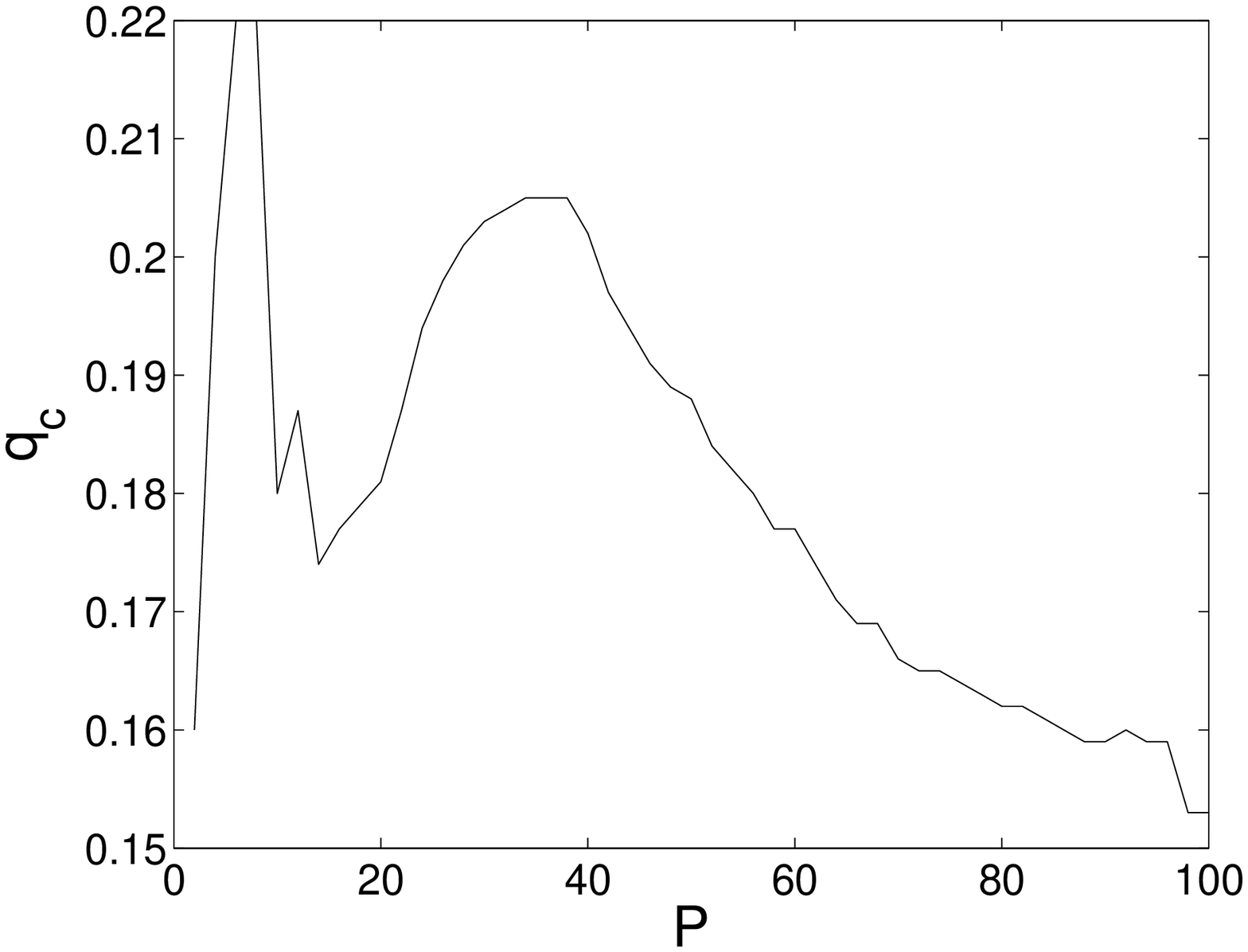}\\
\end{tabular}
\caption{Critical value of the initial thrust $q$ for (a) IS and (b)
OS collisions versus $P$.}%
\label{fig:qc}
\end{center}
\end{figure}

For high values of $P$, the above scenario takes place, except that,
for high values of $q$, \emph{breather creation} is observed. Figure
\ref{fig:example}c shows an example of such a collision. This
behaviour is similar to the soliton creation observed in the saturable
continuous models and will be analyzed in more detail below. The
different regimes in the $(P,q)$ plane are depicted in Figure
\ref{fig:regimes}, for both IS and OS collisions.  Furthermore, Fig.\
\ref{fig:qc} shows the values of the critical value of $q$ separating
merging and reflecting regimes, as a function of the power $P$. It can
be seen that, for most choices of $P$, both values are close. This is
different from the cubic DNLS case \cite{PKMF03} where the critical
values of the OS case is an order of magnitude higher than the ones
for the IS case. The likely explanation is that in the saturable case,
the PN barrier is small (for our choice $\beta=2$, the absolute value
of the barrier is smaller than $0.01$).

For high values of $P$, we have also observed the merging of two
breathers with symmetry breaking, as reported in \cite{PKMF03}. This
symmetry breaking manifests as a movement of the final bound state to
left or to the right accompanied by the appearance of a total lattice
momentum, defined by $p=i\sum_n(\psi_{n+1}\psi^*_n -
\psi^*_{n+1}\psi_n)$.  Since the equation, the initial conditions, and
the boundary conditions are symmetric, this state must be a numerical
artifact, as suggested in \cite{PKMF03}. To test this hypothesis
further, we performed some runs with either (a) increased numerical
accuracy in the numerical integration routines, or (b), the addition
of some very small random noise to the initial conditions.  In case
(a), the onset of symmetry breaking is shifted to longer times,
whereas in case (b), symmetry breaking is observed at shorter times.
These numerical results confirm that symmetry breaking is a numerical
artifact caused by random rounding errors breaking the symmetry of the
problem. However these ``spurious'' results are interesting in their
own right as they suggest that at these higher values of $P$, the
stationary breather formed after collision is more easily set into
motion by a very small perturbation.  To check that the other
phenomena we observe is {\em not} due to numerical artifacts, we have
carried out similar tests on other runs showing different phenomena.
No such sensitivity to random errors of the accuracy of the
integrators is observed.


\section{Breather creation}

\begin{figure}[ht] %fig 3
\begin{center}
\begin{tabular}{cc}
    (a)  & (b)\\
    \includegraphics[width=\middlefig]{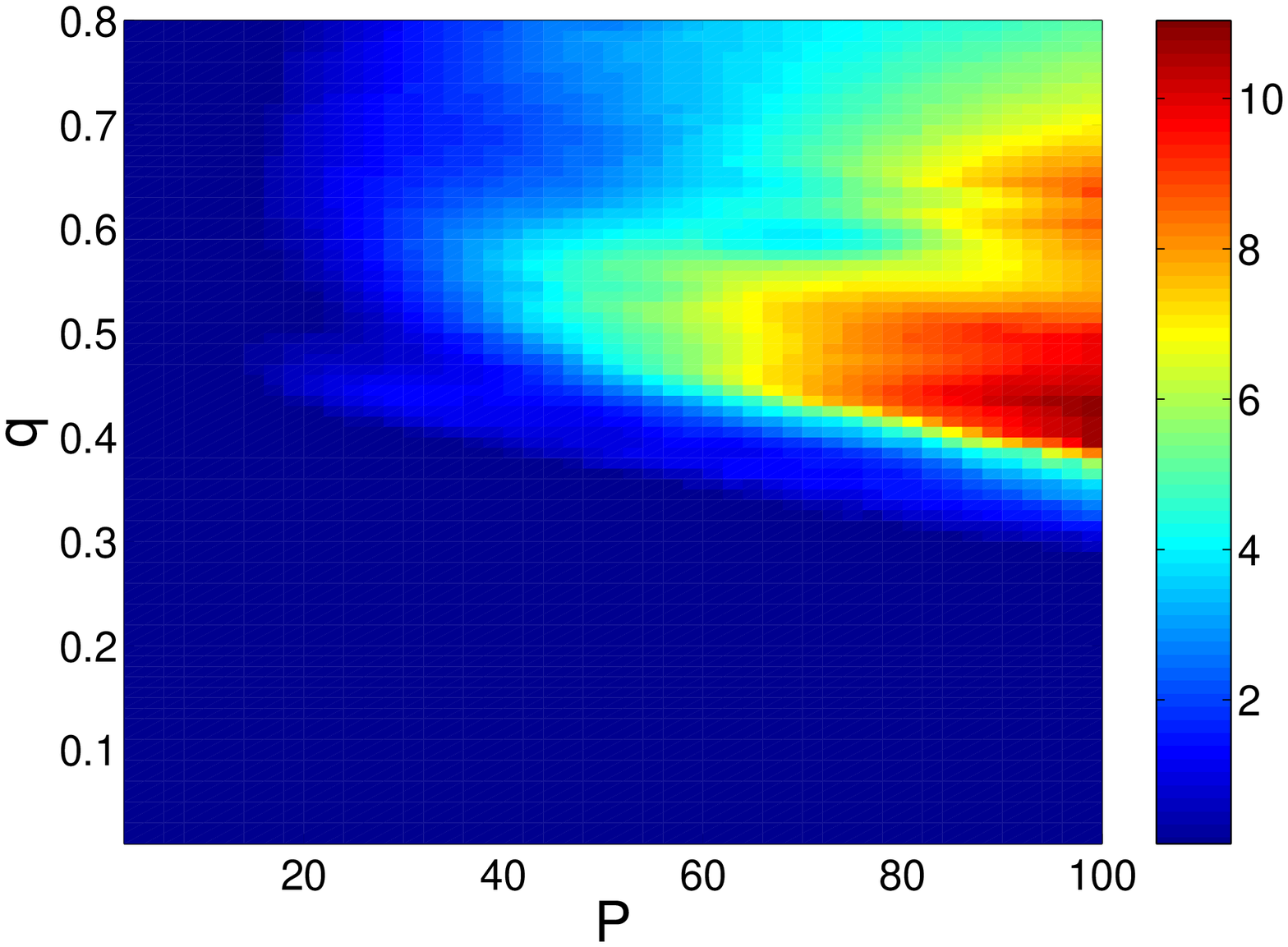} &
    \includegraphics[width=\middlefig]{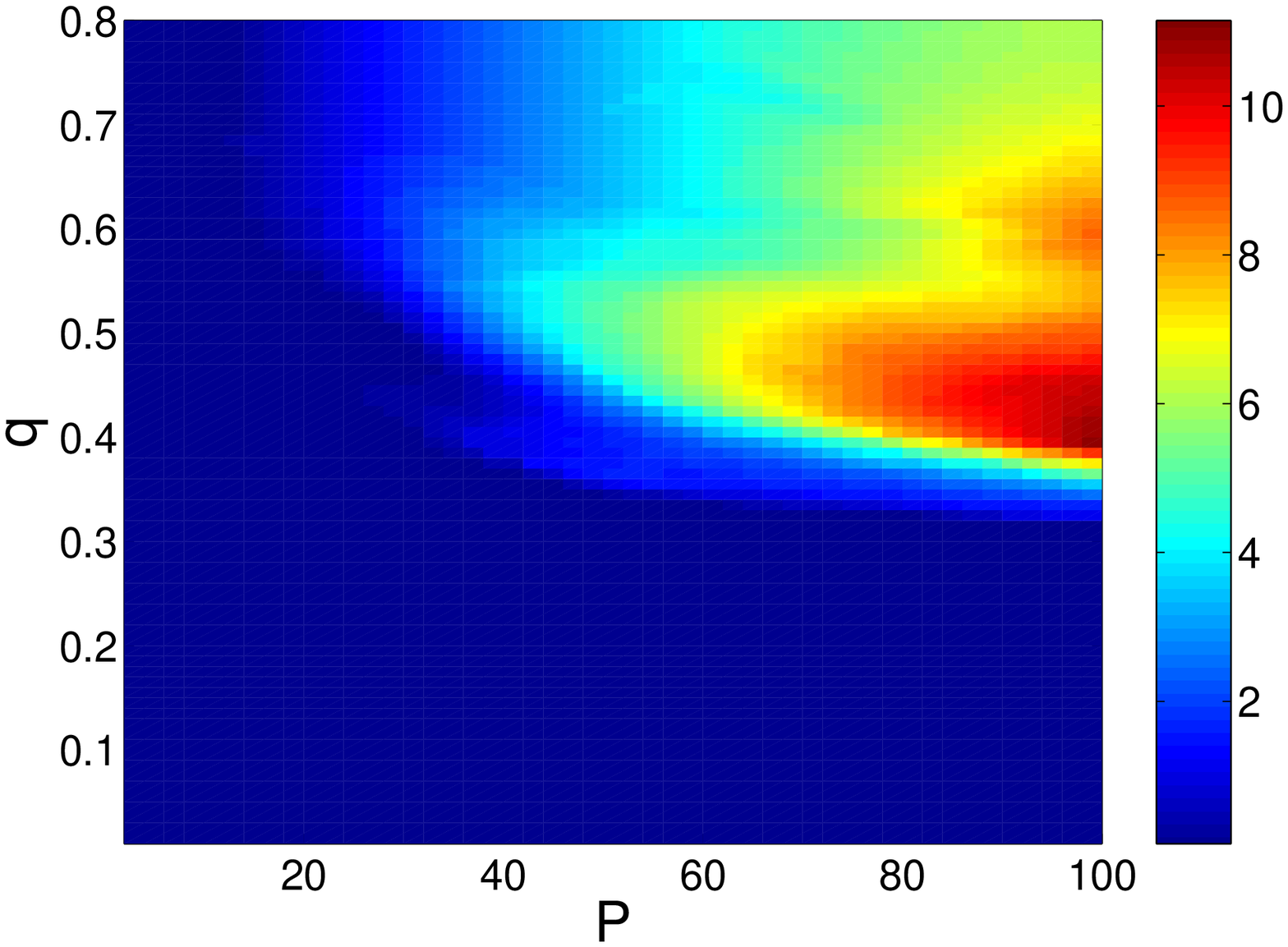}\\
\end{tabular}
\caption{Minimum value of the power density at the collision point
  after collision as a function of $P$ and $q$. Left (right) panel
  corresponds to inter-(on-) site collisions. We have supposed that the
  power trapped in the trapping regime is zero in order to clarify the
  figure.}%
\label{fig:trappow}
\end{center}
\end{figure}

We proceed to analyze the breather creation process, as it is most
noteworthy phenomenon that appears in the saturable case in comparison
to the cubic one. From Figs.~\ref{fig:regimes} and \ref{fig:trappow}
we can conclude that the conditions for breather creation are that $P$
and $q$ are above a threshold value.

\begin{figure}[ht] %fig 3
\begin{center}
\begin{tabular}{cc}
    (a)  & (b) \\
    \includegraphics[width=\middlefig]{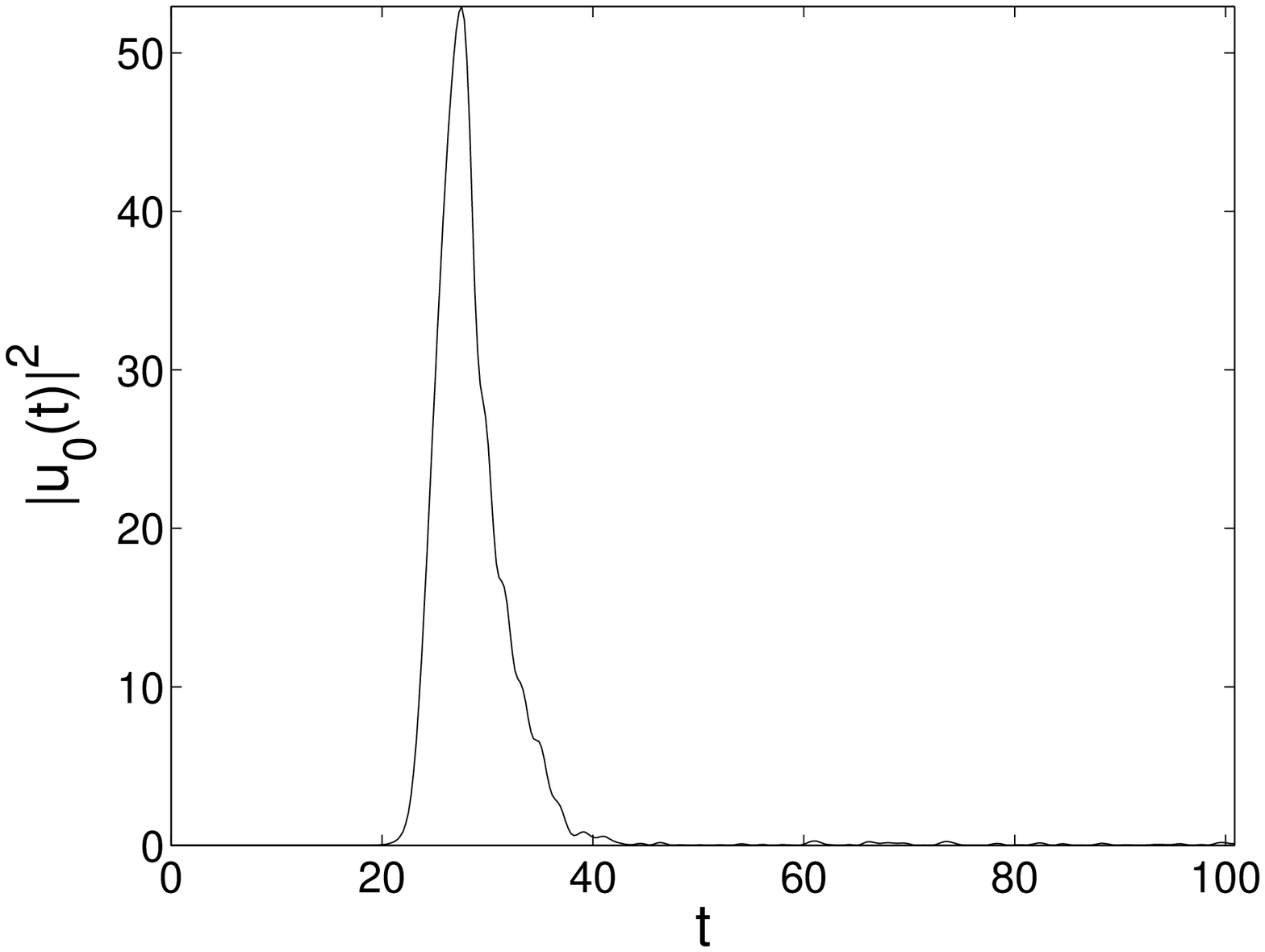} &
    \includegraphics[width=\middlefig]{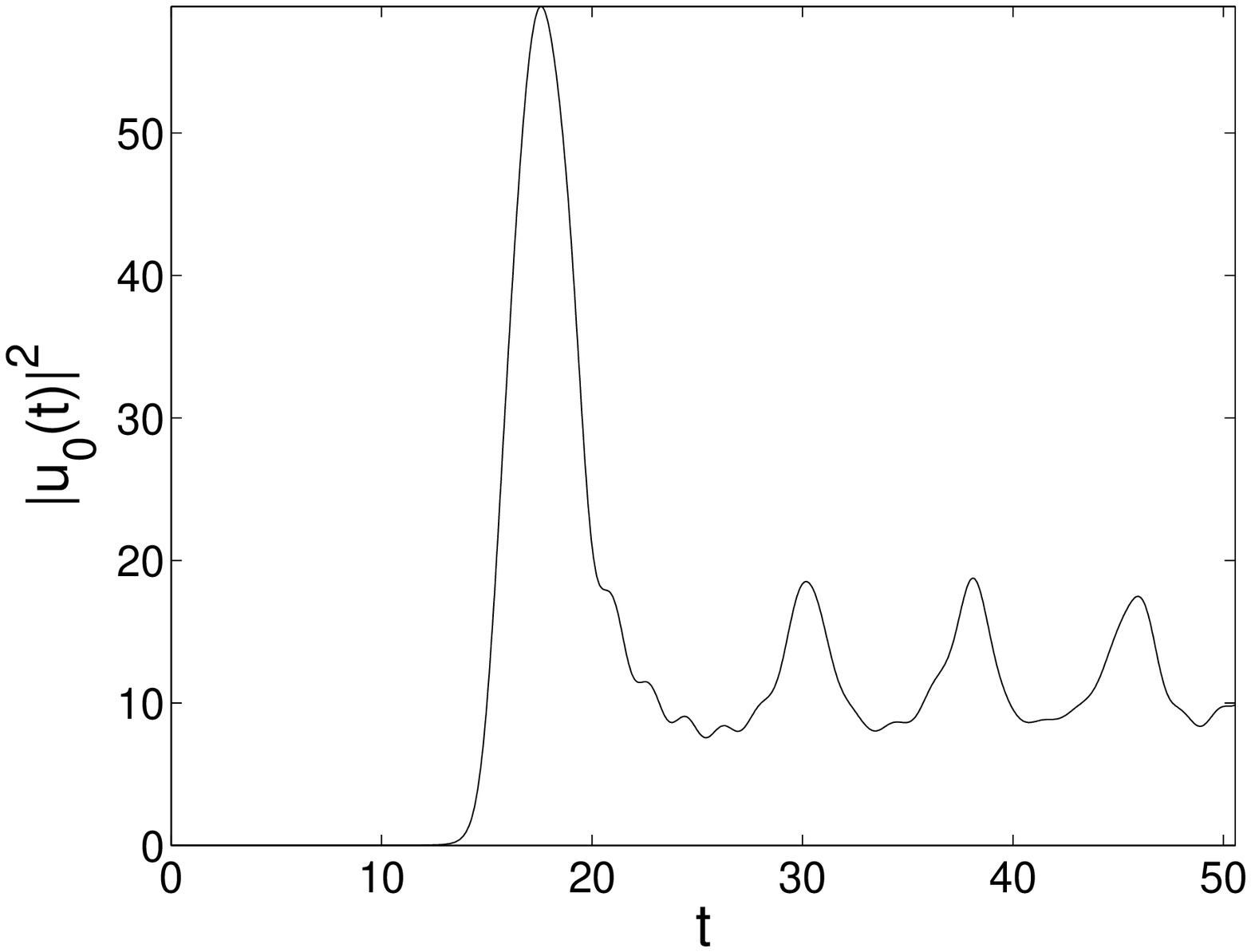}\\
\end{tabular}
\caption{Time evolution of the power density at the collision point
($|u_0|^2$). The left panel corresponds to a reflection case
($q=0.3$, $P=70$) and the right panel to a creation case ($q=0.7$, $P=70$).}%
\label{fig:colpoint}
\end{center}
\end{figure}

\begin{figure}[ht] %fig 3
\begin{center}
\begin{tabular}{cc}
    (a)  & (b) \\
    \includegraphics[width=\middlefig]{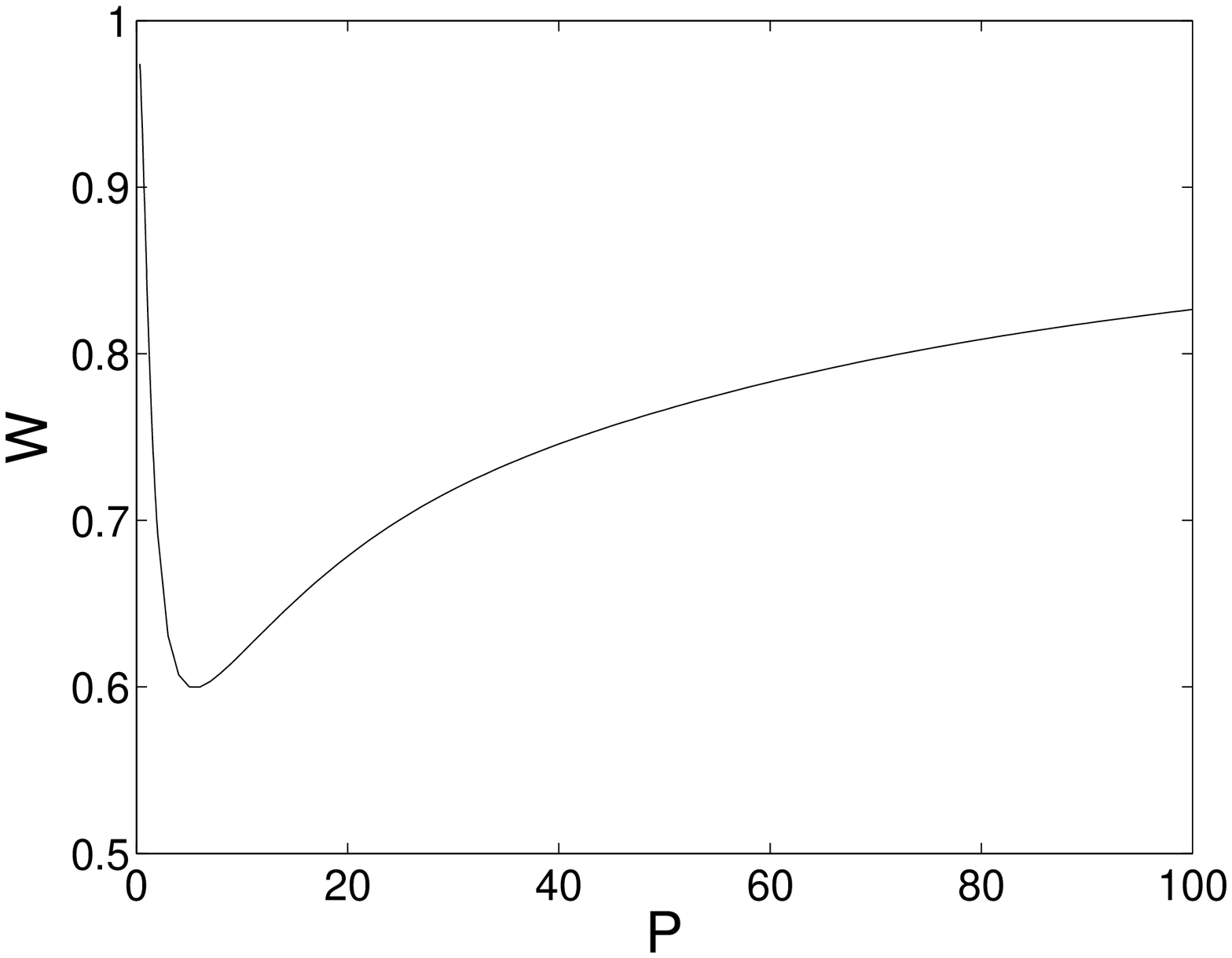} &
    \includegraphics[width=\middlefig]{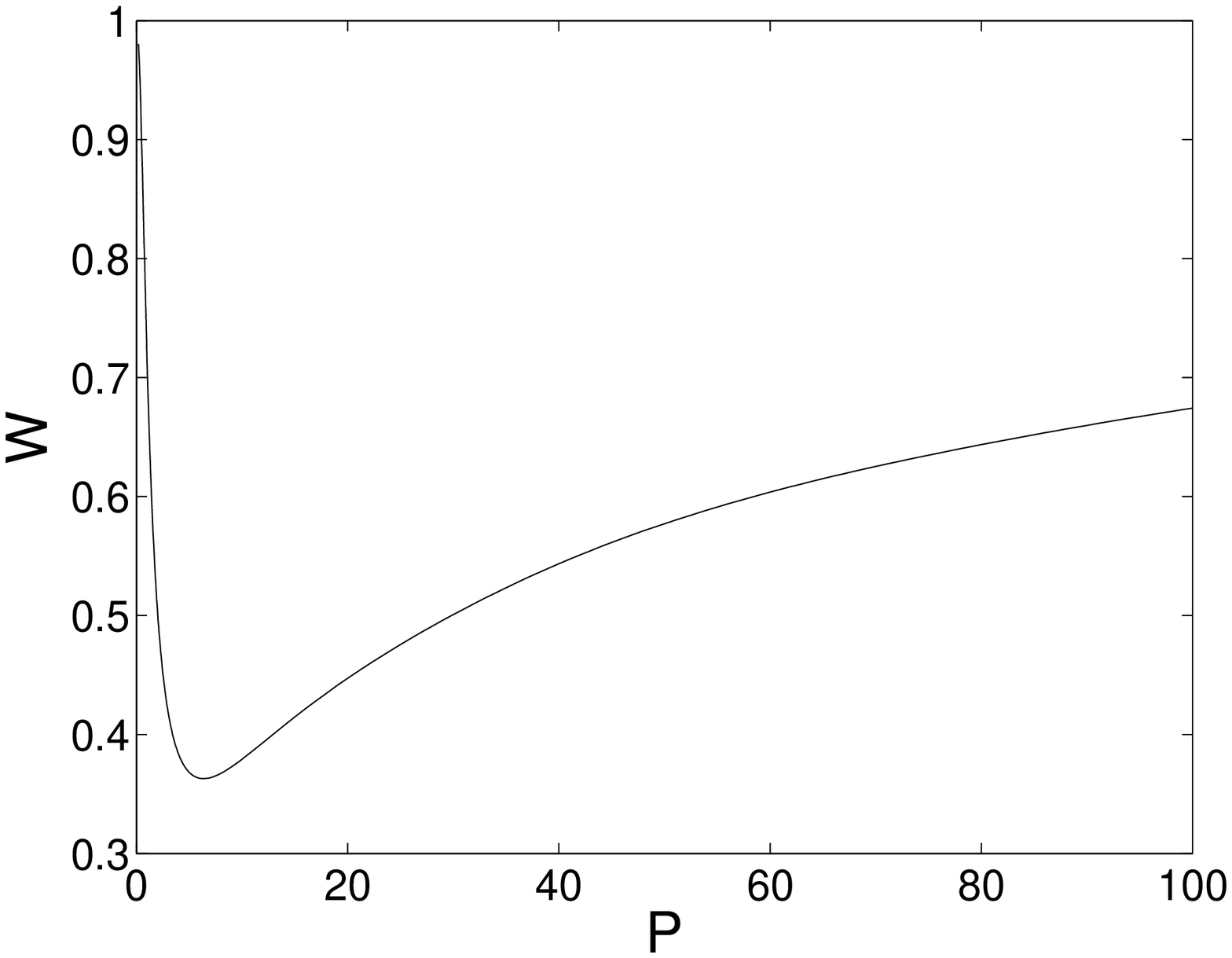}\\
\end{tabular}
\caption{Representation of the ``breather width'' $W$ (defined as
$W=|u_1|^2/|u_0|^2$) versus the power of a stationary site-centred
(left) and a bond-centred (right) breather.}%
\label{fig:width}
\end{center}
\end{figure}

This result can be explained with the aid of Fig.~\ref{fig:colpoint},
where the density power of the collision point, for the cases of
reflection and breather creation, is shown. It can be seen that the
power density oscillates after the collision, and its minimum is zero
for the case of no creation. The minimum power density after the
collision for all the simulations (neglecting the trapping regime) is
represented in Fig.~\ref{fig:trappow}. In consequence, the trapped
power must be above a threshold so that breather creation occurs. It can
be explained by the fact that, for a stationary breather to have a
``saturable'' behaviour, its power should be higher than a threshold
value. This phenomenon is similar to the soliton bistability observed
for SNLS solitons in \cite{Krol2}.  It consists of the existence of a
minimum in the dependence of the soliton width with respect to the
peak intensity.  This dependence is monotonically decreasing in the
cubic NLS, and thus the soliton in a saturable medium has a Kerr
behaviour for small peak intensities (or power). In the discrete case,
as the width is less well-defined, we have considered instead
$W=|u_1|^2/|u_0|^2$, where $n=0$ is chosen as the centre (or peak) of
the breather. Fig.~\ref{fig:width} shows $W$ versus $P$ displaying a
similar behaviour as in the continuous case.

The analysis given above also explain why the results of \cite{PKMF03}
(i.e. only merge and reflection regimes take place) are found for
small values of the power.  We note also that this creation process
may be related to the phenomena of the fission of a coupled
two-breather state into a stationary and a moving breather in the DNLS
equation \cite{ac98}.


\section{Effect of phase in breather collisions}

\begin{figure}[ht] %fig 5
\begin{center}
\begin{tabular}{cc}
    (a)  & (b) \\
    \includegraphics[width=\middlefig]{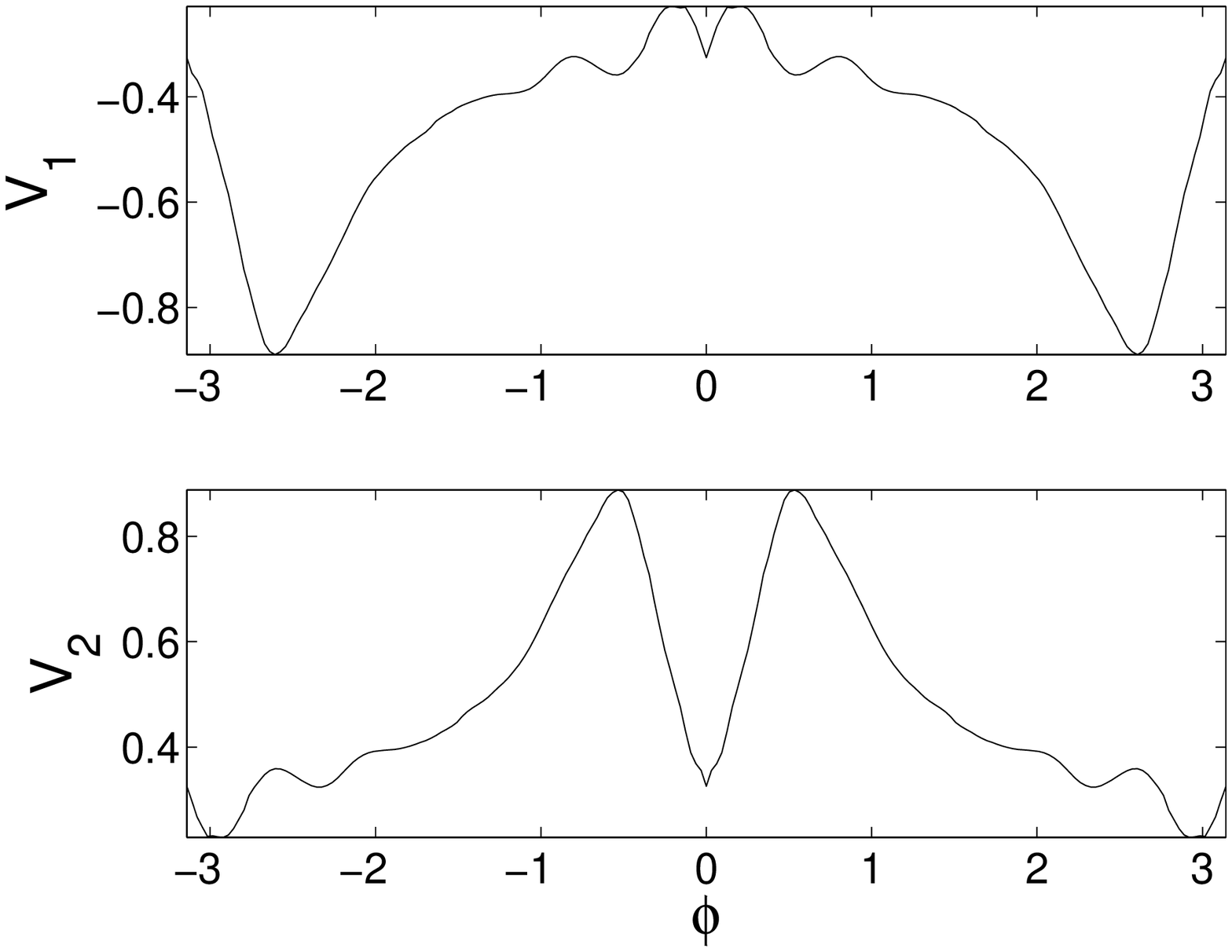} &
    \includegraphics[width=\middlefig]{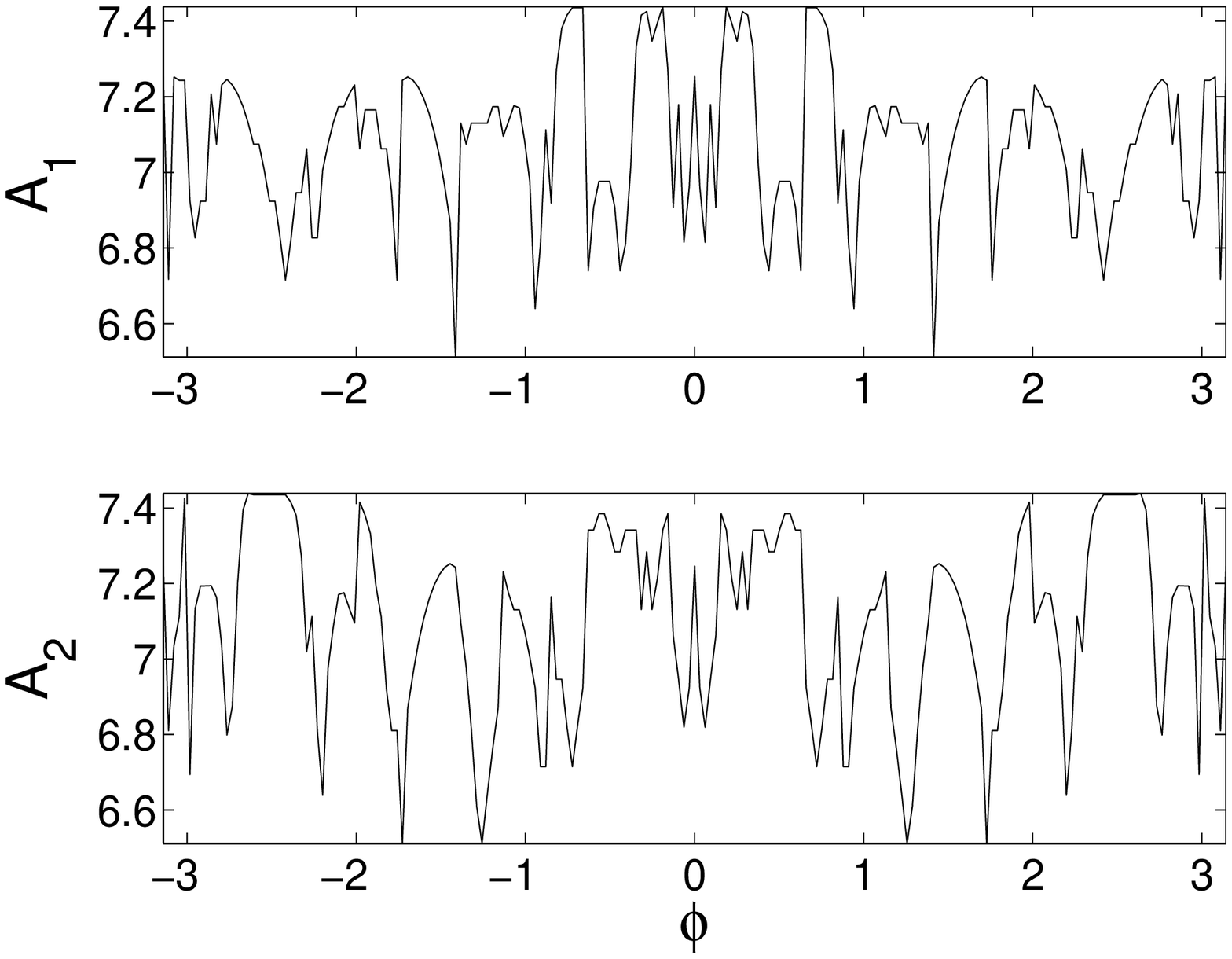}\\
\end{tabular}
\caption{Final velocities (a) and amplitudes (b) with respect to the
phase for breathers with $P=20$ and $q=0.25$ and an OS collision.}%
\label{fig:phase1}
\end{center}
\end{figure}

To complete the paper, we give a brief study of the effect of
considering a phase difference between the breathers, in a similar
fashion to Ref.\ \cite{DKMF03}. This is achieved by introducing a
factor $\exp(i\phi)$ in one of the breathers. Fig.\ \ref{fig:phase1}
shows the final amplitudes $A_{1,2}$ and velocities $V_{1,2}$ of
both breathers as a function of $\phi$.

It appears that the final velocities are smooth functions of $\phi$,
showing a strong phase effect, with the $V_2$ curve following the
$V_1$ curve, phase-shifted by $\pi$.  The $V$ values vary from around
$0.2$ to $0.8$.  The amplitude dependence, on the other hand, is
much smaller but shows a much more irregular behaviour as a function
of $\phi$. Clearly the discreteness of the lattice is featuring
strongly here in this latter case.

\begin{figure}[ht]
\begin{center}
\begin{tabular}{cc}
    (a) & (b) \\
    \includegraphics[width=\middlefig]{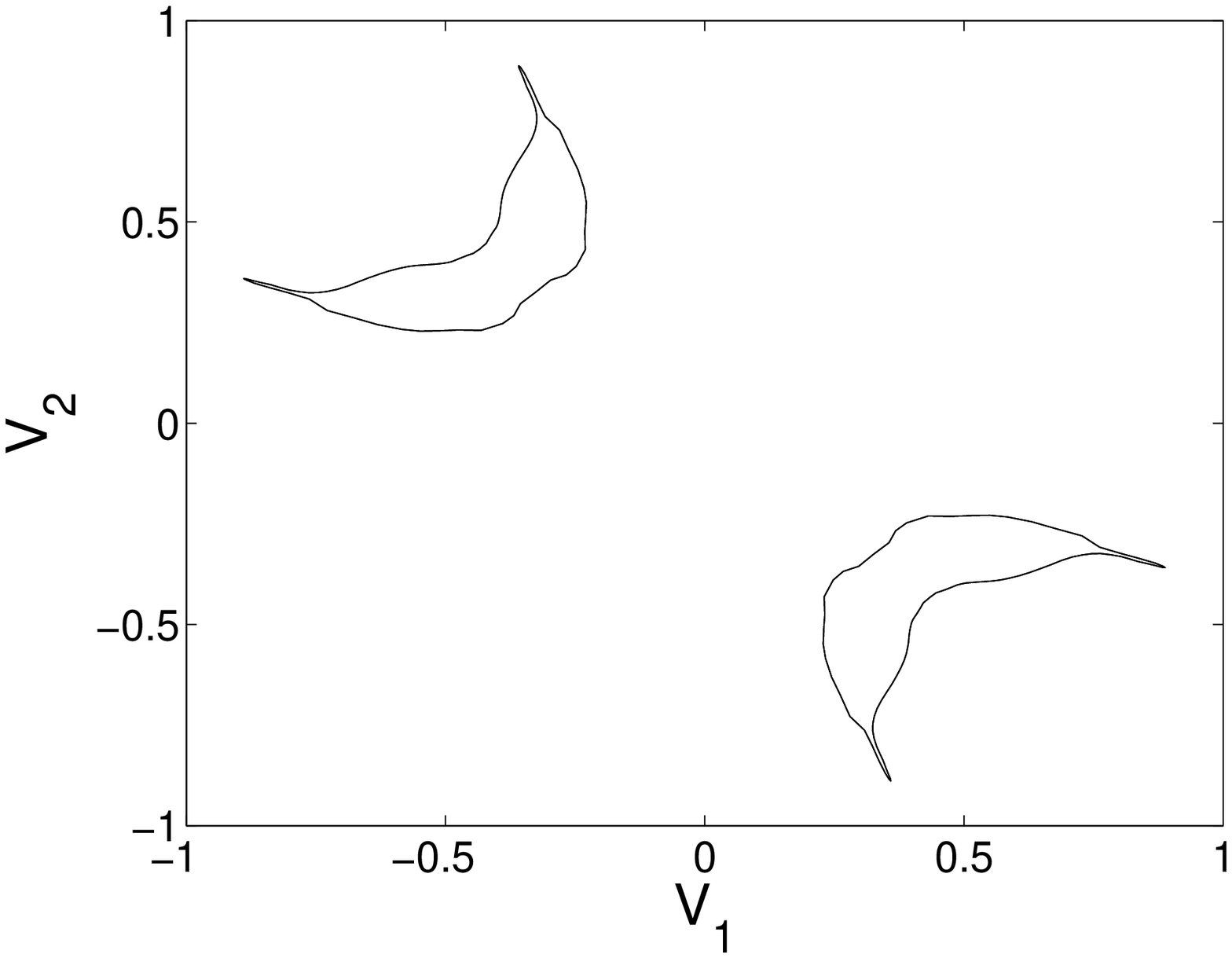} &
    \includegraphics[width=\middlefig]{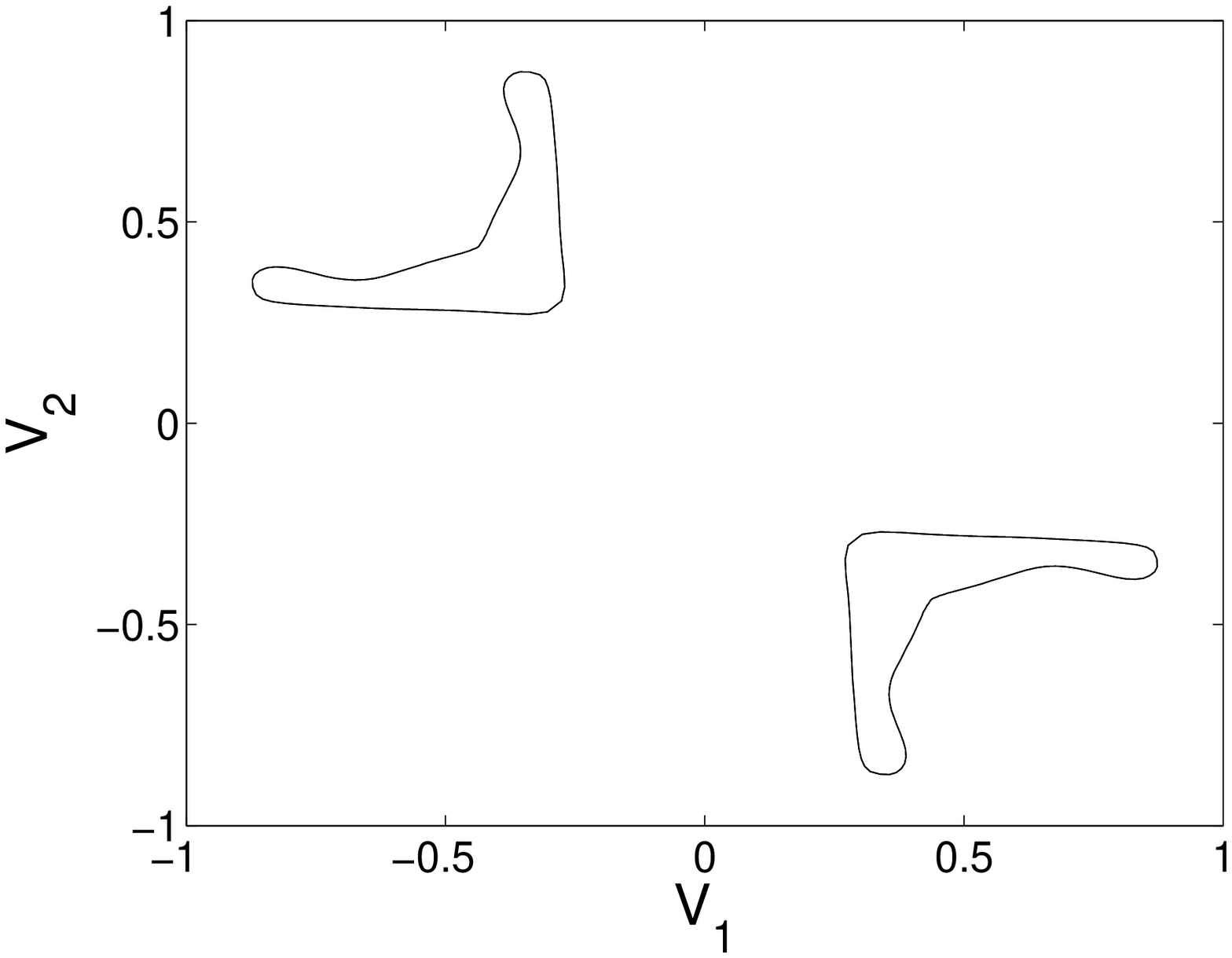}\\
\end{tabular}
\caption{Relation between the final velocities for breathers with
$q=0.25$ and (a) $P=20$; (b) $P=10$.
OS collisions are considered in both cases}%
\label{fig:phase2}
\end{center}
\end{figure}

Finally, in Fig.\ \ref{fig:phase2}, we show the relation between the
outgoing velocities as $\phi$ varies through $2\pi$, analogously to
Fig.\ 3 of \cite{DKMF03}. Here the relatively smooth behaviour over
a large range of $V$ values is clearly shown.

\section{Conclusions}

We have analyzed the collisional behaviour in a saturable DNLS model,
finding close analogies to the continuous NLS equation. Breathers can
merge, reflect or be created (although breather annihilation is not
observed).  The extra power available to breathers in the SDNLS case
results in the new phenomena of breather creation in a discrete
model.  Additionally, the scenario in the saturable DNLS case seems to be
much ``cleaner'' than in the cubic DNLS case on a coarse scale, with a
strong but simpler threshold effect. These facts may be an advantage
in some applications, such as multi-port optical switching. There are
still a number of details in the fine-scale structure which are as yet
unexplained.  These may perhaps be understood through the application
of a future variational study.  It would also be interesting to extend
this study to consider the collision of two non-identical breathers.

\noindent {\bf Acknowledgements} One of us (JC) acknowledges financial
support from the MECD/FEDER project FIS2004-01183.  The other (JCE)
would like to acknowledge the hospitality of the University of Seville
for hosting research visits during which this work was discussed.

\noindent {\bf Notes added in proof}
Since the original version of this paper was written, Khare et al.\
\cite{krss05} have published an exact stationary breather solution for
an equation closely related to (\ref{eq:dyn}), namely
\begin{equation}
  i\dot \psi_n+\frac{\nu|\psi_n|^2}{1+\mu|\psi_n|^2}\psi_n +
  (\psi_{n+1}-2\psi_n+\psi_{n-1})=0. \label{krss}
\end{equation}
 Solutions of (\ref{krss}) can be mapped into solutions of
 (\ref{eq:dyn}) by the (invertible) transformation
\[
\psi_n(t)=\frac1{\sqrt{\mu}}\exp\{i\nu t/\mu\}\, u_n(t), \quad \beta=\nu/\mu,
\]
so the solutions given \cite{krss05} can be mapped into solutions of
(\ref{eq:dyn}).  However the localized stationary breather solution of
\cite{krss05} only exists for $\beta>2$, and hence are not relevant to our
discussions which focus on the $\beta=2$ case.  It would be
interesting to extending the calculations in this paper to other
values of $\beta$ to see if the presence of these stationary solutions
affected the results given here.

%\bibliography{/home/chris/tex/bibs/lattice}
\bibliographystyle{unsrt}

\end{document}